\documentclass[12pt]{iopart}

\usepackage{iopams}
\usepackage{graphicx}
\begin{document}

\title{Charge fluctuations and their effect on conduction in biological ion channels}

\author{D.G. Luchinsky$^{1,4}$, R. Tindjong$^1$, I. Kaufman$^2$,
P.V.E. McClintock$^1$ and R.S. Eisenberg$^3$}

\address{$^1$Department of Physics, Lancaster University, Lancaster, LA1 4YB, UK}
\address{$^2$The Russian Research Institute for Metrological Service,
\\ Gosstandart, Moscow, 119361, Russia}
\address{$^3$Department of Molecular Biophysics and Physiology, Rush Medical College,
\\ 1750 West Harrison, Chicago, IL 60612, USA}
\address{$^4$NASA Ames, Mail stop 269-3, Moffett Field, CA, 94035, USA.}

\ead{r.tindjong@lancaster.ac.uk}

\begin{abstract}
The effect of fluctuations on the conductivity of ion channels is
investigated. It is shown that modulation of the potential barrier
at the selectivity site due to electrostatic amplification of charge
fluctuations at the channel mouth exerts a leading-order effect on
the channel conductivity. A Brownian dynamical model of ion motion
in a channel is derived that takes into account both fluctuations at
the channel mouth and vibrational modes of the wall. The charge
fluctuations are modeled as a short noise flipping the height of the
potential barrier. The wall fluctuations are introduced as a slow
vibrational mode of protein motion that modulates ion conductance
both stochastically and periodically. The model is used to estimate
the contribution of the electrostatic amplification of charge
fluctuations to the conductivity of ion channels.
\end{abstract}

%\pacs{ , }
%\submitto{\JPG}
%\maketitle

\section{Introduction}
\label{s:introduction}

Ion transport through the channels in cellular membranes underlies the
electrical signal transduction and processing by living organisms. Accordingly
ion channels, being natural nanotubes, control a vast range of biological
functions in health and disease. The understanding of their
structure-properties relationship is the subject of intensive, ever-growing,
fundamental and applied research in biology, physics, and
nanotechnology~\cite{Hille:92,Eisenberg:98}. A central problem in studies of
ion permeation through biological membrane channels is to understand how
channels can be both highly selective between alike ions and yet still conduct
millions of ions per second~\cite{Corry:01a}. Indeed, selectivity between ions
of the same charge implies that there exists a deep potential well for
conducting type ions at the selectivity site of the channel. On the other hand
such channels can pass up to 10$^8$ ions per second~\cite{Jordan:05}
corresponding to almost free diffusion.

Significant progress has been made towards an understanding of this problem
over the last few decades. In particular, the molecular structure of the KcsA
potassium channel~\cite{Doyle:98} that discriminates between Na$^{+}$ and
K$^{+}$ was determined by crystallographic analysis. Furthermore, by detecting
the size of the structural fluctuations~\cite{Zhou:01} and conformational
changes~\cite{Dutzler:02b}, it has become possible to provide the experimental
information needed for molecular modelling of the dynamical features of the
observed selectivity and gating~\cite{Corry:06b,Noskov:04a}. In particular, the
minimum radius of the selectivity filter in KcsA is $\sim$0.85{\AA}, which is
to be compared with 1.33{\AA} for the ionic radius of K$^{+}$, suggesting that
flexibility of the filter is coupled to ionic
translocation~\cite{Shrivastava:02a}. It has therefore become apparent that
fluctuations in the channel walls plays a fundamental role in maintaining high
conductivity in highly selective channels (see also Elber~\cite{Elber:95a}).

Another important source of modulation of the electrostatic potential
identified in earlier research~\cite{Heinemann:90a,Eisenberg:04e} relates to
the interaction of the ion in the channel with charge fluctuations in the bath
solutions. The effect of current fluctuations and noise on the channel entrance
rates and on the channel conductivity was also considered
in~\cite{Bezrukov:95a,Adair:03a}. It becomes clear that fluctuations of the
electrostatic potential within ion channels induced by various sources may
provide a key to the solution of the central problems of permeation and
selectivity. Models of such fluctuations have thus become one of the central
topics of research on the permeability of ion channels. It is important to note
that dynamical models of ion motion in the channel can also provide a link
between studies of the permeability of open channels and channels gating.
Notwithstanding recent advances, theoretical modeling of the dynamical features
of ion channels is still in its infancy. In particular, little is yet known
about the relative importance of the different dynamical mechanisms and sources
of fluctuations in the ion channels.

In our earlier work we have started to develop a dynamical
model~\cite{Tindjong:04,Tindjong:05,Tindjong:06,Tindjong:07,Luchinsky:07,Kaufman:07}
of ionic conductivity through open channels. It takes into account the coupling
of ion motion to vibrations of the wall~\cite{Tindjong:04,Tindjong:05} and to
charge fluctuations at the channel mouth~\cite{Tindjong:07,Luchinsky:07}. Our
goal is to derive a self-consistent model that allows for analytical estimation
of the potential barrier at the selectivity site and for the effects of
fluctuations on the conductivity of the channels. The starting point of our
approach is a self-consistent quasi-analytical solution of the Poisson and
Nernst-Planck equations in the channel, and in the bulk~\cite{Kaufman:07},
allowing for accurate estimation of the current-voltage characteristics of ion
channels~\cite{Luchinsky:08b} (see also
~\cite{Chen:92,Eisenberg:95b,Eisenberg:98a,Eisenberg:99a,Eisenberg:99}). The
electrostatic channel potentials resulting from these estimates can be further
used to estimate relative contribution to the channel conductivity from the
different sources of fluctuations.

In this paper we introduce a model of ion permeation that takes into
account dynamical effect of the charge fluctuations through the
resultant shot noise, and we demonstrate that the latter has a
leading-order effect on the transition probabilities. We show that
the charge fluctuations at the channel mouth can be modeled as a
flipping of the electrostatic potential at the selectivity site,
which fluctuates between two maximum values at a rate corresponding
to the random arrivals of ions at the channel mouth. The developed
theoretical framework will allow us in the future to include into
the model the modulation of the potential at the selectivity site
due to hydration effects inside the channel.

A model of 3D Brownian dynamics simulation of ions in the bulk and
inside the channel is described in Sec.~\ref{s:BD}. Using results of
the 3D simulations in the bulk we present in Sec.~\ref{s:Reduce
model} a reduced model of an ion moving in the channel and
interacting with the wall vibrational modes and with charge
fluctuations at the channel entrance. The model uses the channel
potential derived from a self-consistent solution of the Poisson
equation and the flipping rates of the potential barrier obtained
from simulations of Brownian motion of ions in the bulk. In
Sec.~\ref{s:crossing_time}, we estimate analytically the mean first
passage time of the channel. These estimates are based on the
assumption that barrier-crossing and barrier-fluctuations are
correlated for a general form of the potential
barrier~\cite{Doering:93a}. The mean first passage time is
calculated as a weighted sum of the escape time $\tau_{-}$ over the
low barrier $\Delta E_{0}$, and the escape time $\tau_{+}$ over the
high barrier $\Delta E_{1}$. Conclusions are drawn and future
directions of research are outlined in Sec.~\ref{s:conclusions}.

\section{Brownian dynamics simulations in the bulk and inside the channel}
\label{s:BD}

\begin{figure}[!t]
   \begin{center}
   \includegraphics[height=5cm, width=6.5cm]{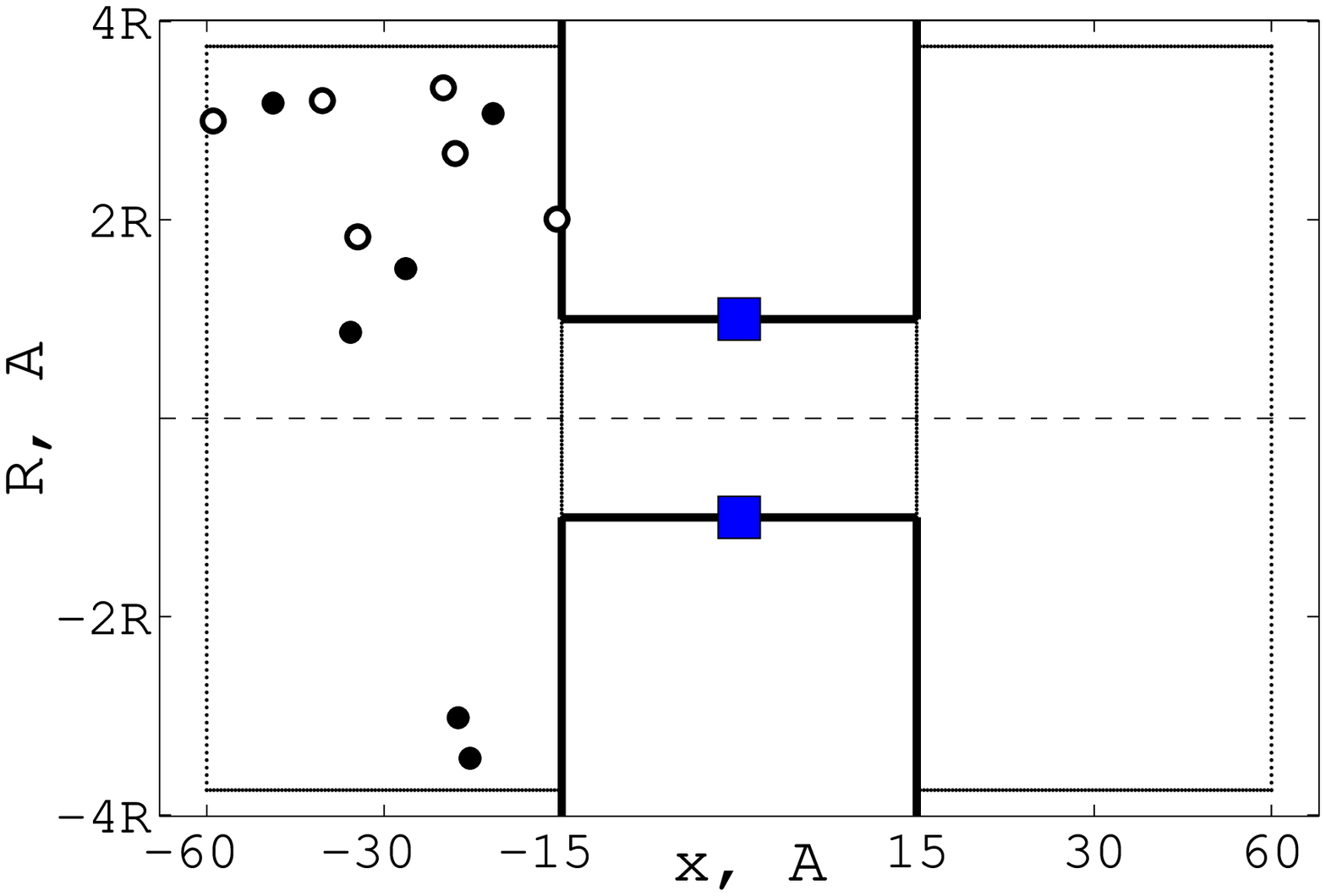}
   \includegraphics[height=5cm, width=6.5cm]{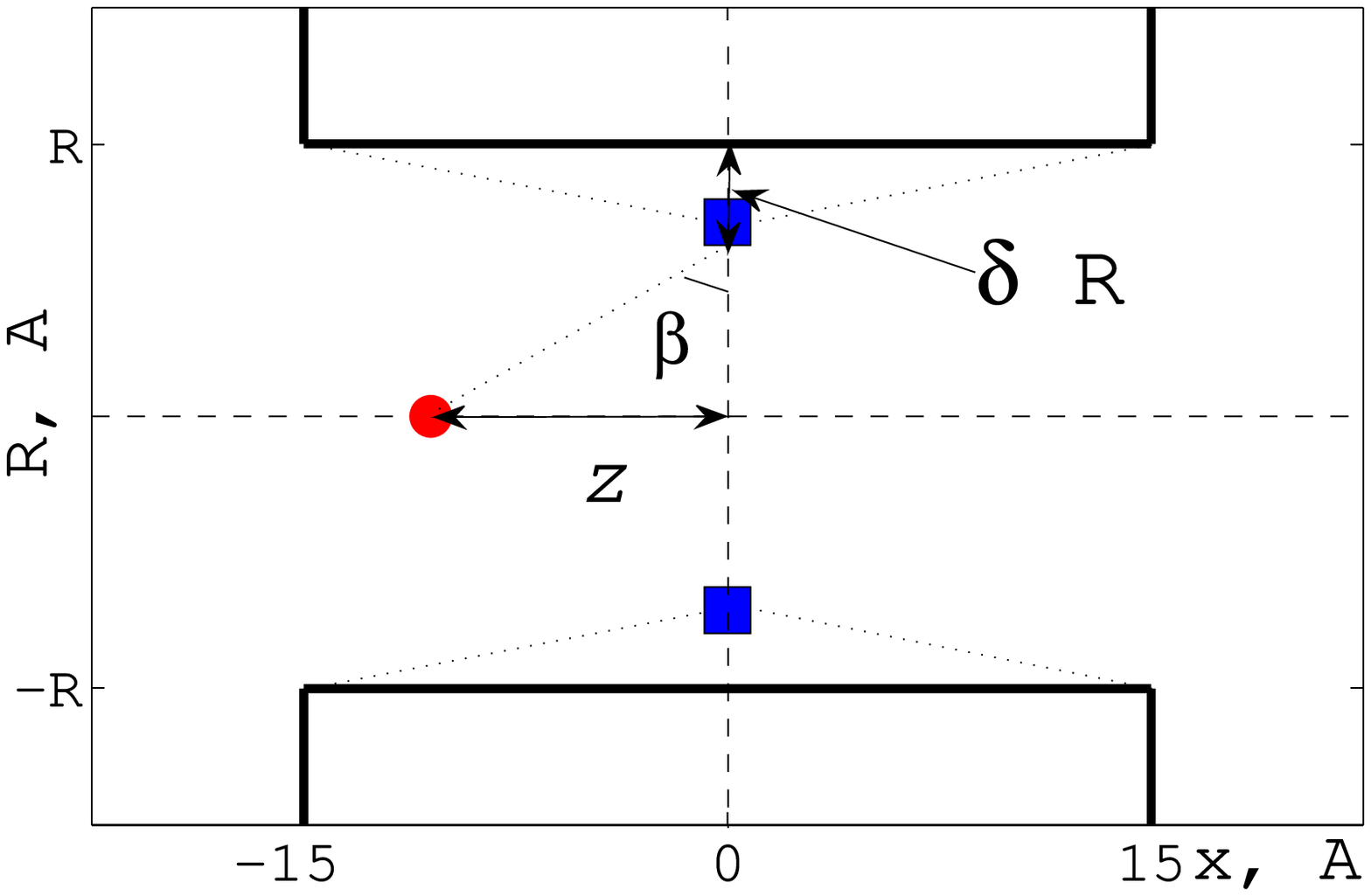}
   \end{center}
\caption{{\bf Left:} Sketch of the model. The middle block
represents the protein, through which there is a cylindrical hole
approximating the open ion channel. The moving charged segment of
the protein wall, assumed elastically bound to the wall, is shown by
the filled squares. Negative ions are shown by the filled circles,
and positive ions by the open circles. {\bf Right:} Sketch of an ion
moving along the channel axis. The conducting ion is shown by red
circle. The difference between ion coordinate $x$ and location of
the selectivity site $x_{ss}$ is $z$. The vertical displacement of
the selectivity site is $\delta R$. The angle between a line
connecting the ion with the selectivity site and the vertical axis
is $\beta$.} \label{fig:moving_wall}
  \end{figure}

We consider the following Brownian dynamical model of the ion permeation (see
sketch in the Fig.~\ref{fig:moving_wall} (left)). The system is made of three
compartments of equal size. The middle block constitutes the protein through
which there is a cylindrical hole approximating the open channel. To model the
ion's coupling to the vibrational modes of the channel, we introduce a moving
segment of the protein wall that is elastically bound to the wall. The moving
segment is charged and represents the selectivity site. The dynamics of the
ions in the bath and channel, and of the moving segment, are modeled using
Brownian Dynamics (BD) simulations, see eqs.\
(\ref{eq:langevin_1})-(\ref{eq:wall}) coupled to the Poisson equation
(\ref{eq:poisson_1}).
\begin{eqnarray}
\label{eq:poisson_1}
-\nabla\cdot(\varepsilon(\vec{r})\nabla\phi(\vec{r}))=\rho(\vec{r}),
\end{eqnarray}
\begin{eqnarray}
\label{eq:langevin_1}
 m_{i}\ddot{\vec{x}}_{i}+m_{i}\gamma_{i}\dot{\vec{x}}_{i}=\vec{F}_{C,i}+\vec{F}_{sr,i}+\vec{F}_{H,i}
+\sqrt{2m_{i}\gamma_{i}k_{B}T}\vec{\xi}_{i}(t),
\end{eqnarray}
\begin{eqnarray}
\label{eq:langevin_2}
 m_{j}\ddot{\vec{x}}_{j}+m_{j}\gamma_{j}\dot{\vec{x}}_{j}=\vec{F}_{C,j}+\vec{F}_{sr,j}+\vec{F}_{H,j}
+\sqrt{2m_{j}\gamma_{j}k_{B}T}\vec{\xi}_{j}(t),
\end{eqnarray}
\begin{eqnarray}
 \label{eq:ion_in_channel}
 m\ddot x+m\gamma\dot x=F_{C}
+F_{ch}+F_{iw,i}\sin(\beta)+\sqrt{2m\gamma k_{B}T}\xi(t),
\end{eqnarray}
\begin{eqnarray}
 \label{eq:wall}
 M\delta\ddot{R}+M\Gamma\delta\dot{R}+K\delta R=
F_{iw,M}\cos(\beta)+\sqrt{2M\Gamma k_{B}T}\nu(t)
\end{eqnarray}

\noindent Here $\vec{r}_{ij}=\vec{x}_{i}-\vec{x}_{j}$, $r_{ij}=|\vec{r}_{ij}|$,
$m_i$,~$\vec{x}_{i}$~and $q_{i}$, $m_{i}\gamma_{i}$ and $\sqrt {2m_i \gamma _i
k_BT}\, \vec \xi _i (t)$ are the mass, position, charge, friction coefficient
and the stochastic force of the $i$th ion. The distance between ions $i$ and
$j$ is $r_{ij}$. In this work, for simplicity, we restrict the analysis to two
types of ions: the index $i$ will correspond to Na$^+$, while index $j$
corresponds to Cl$^{-1}$. The motion of the charged residual of mass $M$ at the
selectivity site is characterized by the displacement $\delta R$ in the
vertical direction from the equilibrium position $R$ and an elastic force
$\propto K\delta R$. Note that in general values of the effective mass and
diffusion constant of the ion moving within the channel may deviate from the
corresponding values in bulk due to the nontrivial effect of hydration in the
channel. Coupling between the motion of the ion in the channel and the normal
mode of the wall oscillations is accounted for by the term $F_{iw}$
corresponding to the Coulomb interaction between ion and charge at the
selectivity site. Keeping only the axial component of the force for the ion
motion in the channel, and the radial component of $F_{iw}$ for the oscillating
wall, we have
\[F_{iw,i}=\frac{Qq_{j}}{4\pi\epsilon\epsilon_{0}d^2}\frac{z}{d}, \qquad
F_{iw,M}=\frac{Qq_{j}}{4\pi\epsilon\epsilon_{0}d^2}\frac{\delta
R}{d}.\] Here $d=\sqrt{\left((R+\delta R)^{2}+z^{2} \right)}$,
$z=x-x_{ss}$ where $x$ is the coordinate of the ion in the channel
along the channel axis, and $x_{ss}$ is the location of the
selectivity site. An additional coupling of the ion motion to
fluctuations of the channel wall is effected via modulation of the
channel potential by the moving wall. This is taken into account
through the term $F_{ch}$ on the rhs of eq.\
(\ref{eq:ion_in_channel}). The ions in the bulk (see eqs.
(\ref{eq:langevin_1}), (\ref{eq:langevin_2})) do not feel the
channel potential. Instead, their motion is governed by the Coulomb
interaction $F_{C}$, by the short-range interaction $F_{sr}$, and by
hydration forces $F_{H}$ (see \cite{Corry:01a})
\[
  \begin{array}{ll}
&\vec{F}_{C,i} =
\sum_{j=1}^{N}\frac{q_{i}q_{j}}{4\pi\varepsilon\varepsilon_{0}r_{ij}^{2}}\frac{\vec{r}_{ij}}{r_{ij}},\qquad
\vec{F}_{sr,i} =
\sum_{j=1}^{N}\frac{9U_{0}R_{c}^{9}}{r_{ij}^{10}}\frac{\vec{r}_{ij}}{r_{ij}},\\
&\vec{F}_{H,i} =
\sum_{j=1}^{N}AU_{0}\exp\bigg(\frac{R_{ij}-r_{ij}}{a_{e}}\bigg)
\sin\bigg(2\pi\frac{R_{ij}-r_{ij}}{a_{w}}-\alpha\bigg)\bigg]\frac{\vec{r}_{ij}}{r_{ij}},
  \end{array}
\]
where $A=\sqrt{1+(\frac{a_{w}}{2\pi a_{e}})^{2}}$ and
$\alpha=\arctan(\frac{a_{w}}{2\pi a_{e}})$.

In addition the effect of the surroundings is taken into account by
an average frictional force with frictional coefficient
$m_{i}\gamma_{i}$ and a stochastic force
$\sqrt{2m_{i}\gamma_{i}k_{B}T}\vec{\xi}_{i}(t)$. The addition of the
pairwise repulsive $1/r^{9}$ soft-core interaction potential ensures
that ions of opposite charge, attracted by the inter-ion Coulomb
force, do not collide and neutralize each other. $U_{0}$ and $R_{c}$
are respectively the overall strength of the potential and the
contact distance between ion pairs. The oscillating part is added to
the potential and takes explicitly into account the internuclear
separation for the two solvents, where $a_{w}$ is the oscillation
length, $a_{e}$ is the exponential drop parameter, and $R_{ij}$ is
the origin of the hydration force which is shifted from $R_{c}$ by
$+0.2$ {\AA} for like ions and by $-0.2$ {\AA}
otherwise~\cite{Chung:00a}. $F_{ch}$ is the dielectric force in the
channel, obtained by solving Poisson's equation numerically using
finite volume methods (FVM)~\cite{Ferziger:96}. We use the Langevin
equation to model the collective motion of the atoms forming the
channel protein charged ring located at the selectivity filter. In
this way, our analysis is based on the assumption that the movement
of structural domains of the channel protein may be described as the
motions of independent, elastically bound Brownian
particles~\cite{Lauger:85}. We have included the damping term
$M\Gamma\delta \dot R$ and the corresponding random force
$\sqrt{2M\Gamma k_BT}\nu(t)$, whose amplitude is related to the
damping constant via the fluctuation-dissipation theorem. The
function $\nu(t)$ is a Gaussian white noise. $Q$ is the total fixed
charge on the flexible ring interacting with an ion of charge
$q_{j}$ on the channel axis $z$. $R$ is the channel radius, $\delta
R$ is a small variation of the channel radius and $K$ is the elastic
constant of the channel protein. In the particular case of the
Gramicidin A (GA) channel, the value of the elastic constant is
estimated by calculating the root-mean-square deviation (RMSD) of
the backbone forming its central part, together with the single ion
potential of a Na$^{+}$ ion as it traverses the channel. Using data
from~\cite{Kurnikova:03}, we obtain an elastic constant of
$\sim1.6577$~N/m for a maximum RMSD of $1$~\AA. The GA channel
molecular weight $M$ is about
$4kDa=6.64\times10^{-24}$~kg~\cite{Andersen:05}. The diffusion
constant of the protein in the membrane is between $10^{-14}$ and
$10^{-16}$~m$^2$/s~\cite{Tuszynski:03}.

The parameters of the ion-ion interaction are presented in
Table.~\ref{tab:hydration parameters}.
%---------------------
\begin{table}[!ht]
\caption{\label{tab:hydration parameters}Parameters used in the calculation of the short range
ion-ion interaction with hydration}
\begin{center}
\begin{tabular}{|l|r|r|r||l|l|}
\hline
Ions & {\it U$_{0}$}[k$_{B}$T] & {\it R$_{c}$}[\AA] & {\it R}[\AA] & {\it a$_{w}$}[\AA] & {\it a$_{e}$}[\AA]\\
\hline
Na-Na & 0.5 & 3.50 & 3.7  &   &   \\
Na-Cl & 8.5 & 2.76 & 2.53 & 2.76 & 1.4 \\
Cl-Cl & 1.4 & 5.22 & 5.42 &   &   \\
\hline
\end{tabular}
\end{center}
\end{table}
%----------------

Other parameters used in simulations are:

Dielectric constants: $\varepsilon_2=80$, $\varepsilon_1=2$;

Masses (in kg): $m_{Na} = 3.8\times10^{-26}$,
$m_{Cl}=5.9\times10^{-26}$;

Diffusion coefficients (in m$^{2}$s$^{-1}$):
$D_{Na}=1.33\times10^{-9}$, $D_{Cl}=2.03\times10^{-9}$,

(Note that D is related to the friction coefficient via
$D=\frac{k_{B}T}{m\gamma}$);

Ion radii (in \AA): $r_{Na}=0.95$, $r_{Cl}=1.81$;

Temperature: $T = 298$ K.

\section{Reduced model for ion channel conduction}
\label{s:Reduce model}

To derive the reduced model we notice that Eqs.\
(\ref{eq:poisson_1})-(\ref{eq:wall}) correspond to a many-body
problem with widely-varying timescales, ranging from ps (ion
fluctuations) to $\mu$s (wall vibrations). We further assume that
the channel is occupied most of the time by only one ion, and that
the transition rate of ions through the channel is mainly determined
by escape over the potential barrier at the selectivity site. Then
the effect of the many-body ion dynamics in the bulk on the ion
motion in the channel is twofold: (i) a delivery of the ions to the
channel mouth and (ii) modulation of the channel potential by the
charge fluctuations at the channel mouth. Under these
physiologically plausible assumptions one can separate the ion
motion in the channel from the many-body ion dynamics in the bulk.
The resulting equations of ion dynamics in the channel coupled to
the wall fluctuations can be written as follows
\begin{eqnarray}
\label{eq:overdamped ion}
 &&m\gamma\dot{x}=-\frac{dV(x,t)}{dx} +
\sqrt{2m\gamma k_{B}T}\,\xi(t), \\
\label{eq:damped wall}
 &&M\delta\ddot{R}+M\Gamma\delta\dot{R}+K\delta R= F_{iw,M}\cos(\beta)+\sqrt{2M\Gamma
 k_{B}T}\nu(t).
\end{eqnarray}
Note that the reduced motion of the conducting ion is overdamped
while the wall fluctuations are damped. Damped vibrational mode
models the relatively slow (on a time scale of ns) motion of the
protein of the channel walls that was
suggested~\cite{Skerra:87a,Heinemann:90a} to be essential for the
ion transport process.

\begin{figure}[!h]
\begin{center}
%   %\includegraphics[width=3.in]{.eps}
   \includegraphics[height=5.5cm, width=7cm]{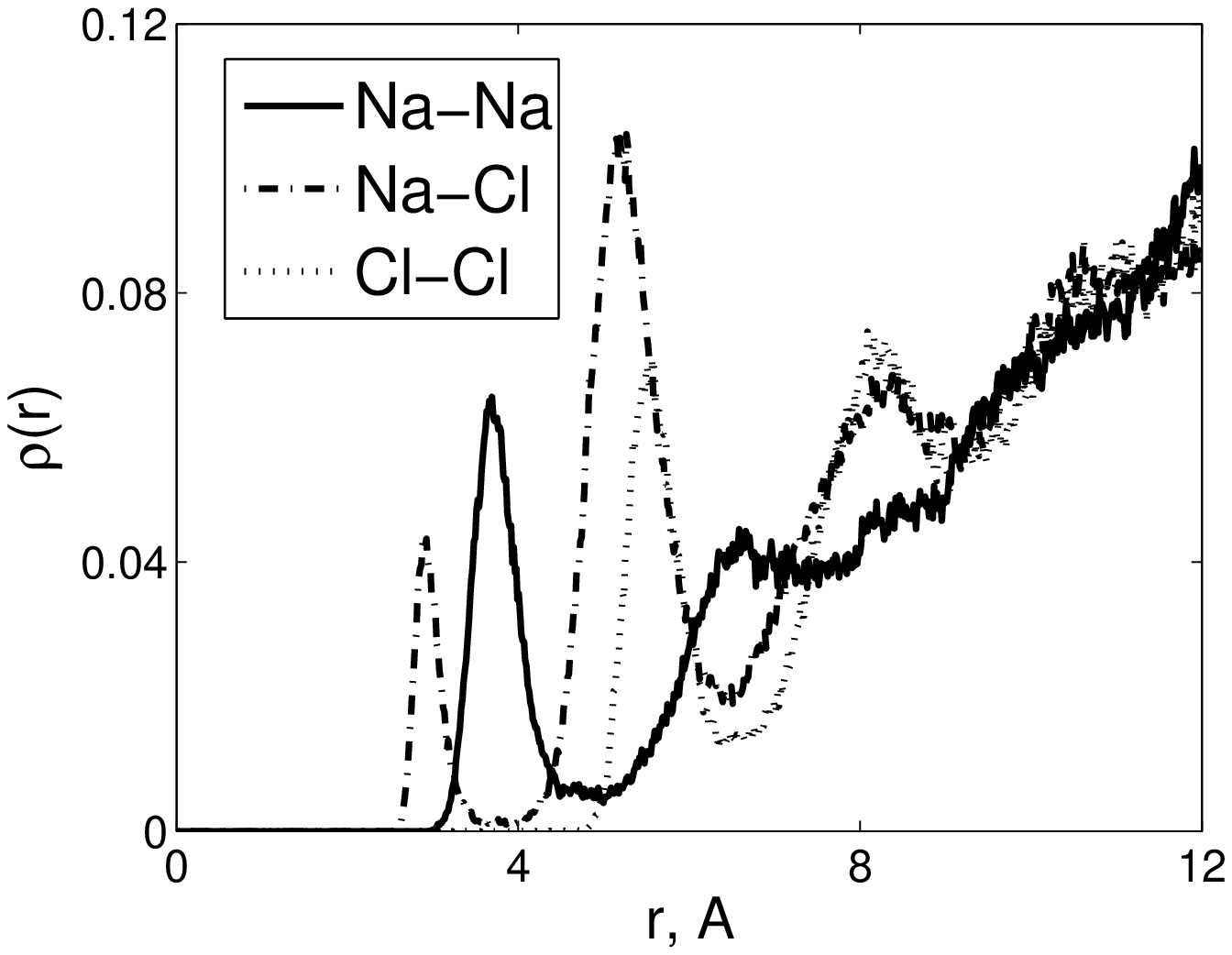}
   \includegraphics[height=5.5cm, width=7cm]{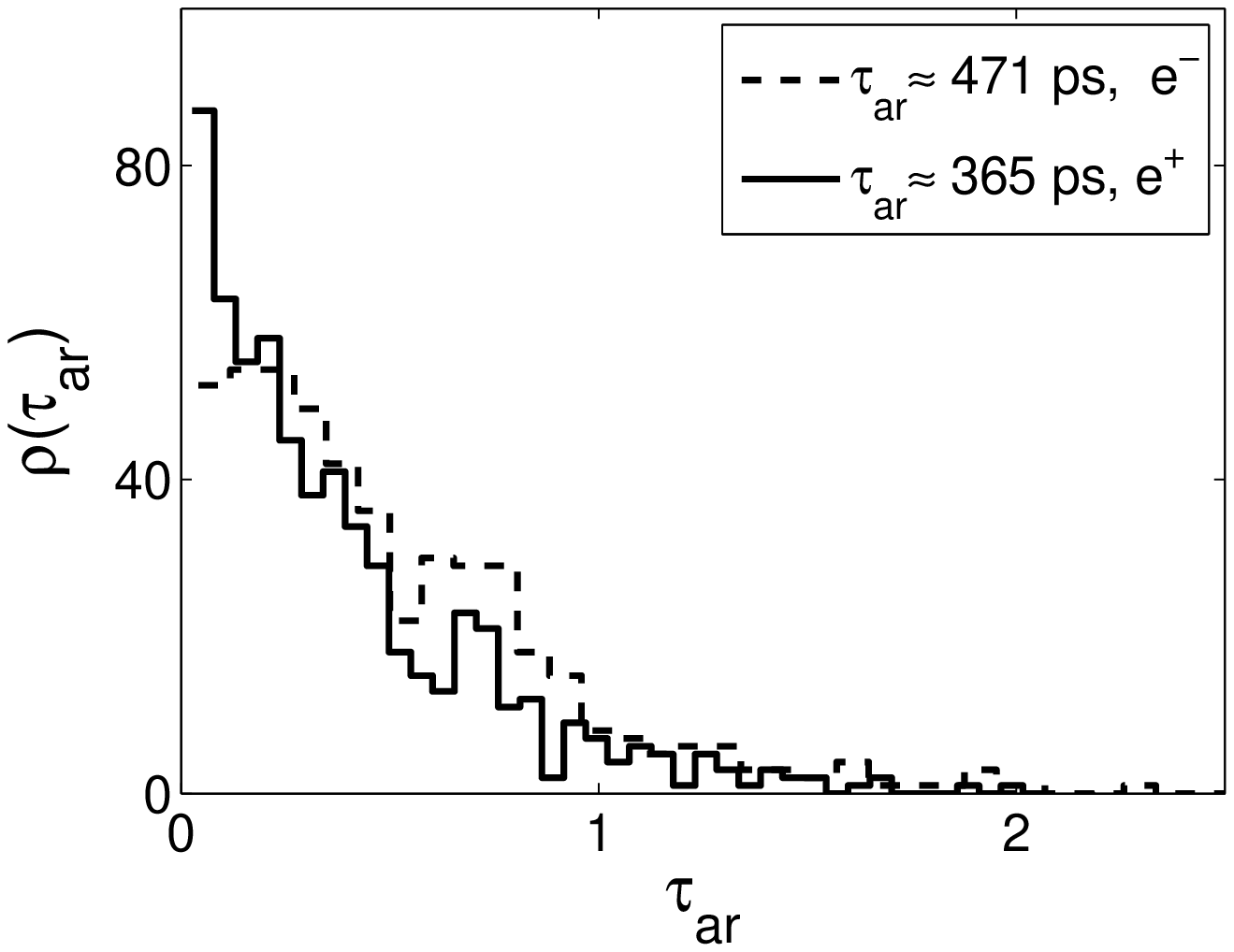}
   \end{center}
\caption{Simulations of Brownian dynamics in the bulk liquid. {\bf
Left:} The number of ions as a function of the distance between ions
of different types: C=400\,mMol; Box size: 40\,A. Forces included in
the simulations were: (i) Coulomb interaction; (ii) short range
repulsion; (iii) hydration. {\bf Right:} The arrival time
distribution for positives(solid line) and negative(dashed line) for
cylindrical channel of radius r=6\AA } \label{fig:distribution}
\end{figure}

The potential $V(x,t)$ in our approximation has three main
contributions: (i) the potential of Coulomb interaction with ions in
the bulk solution $V_C$; (ii) the electrostatic potential induced by
interaction with the channel protein $V_{ch}$; (iii) the potential
of Coulomb interaction with the wall charge at the selectivity site.
By an averaging procedure, the effect of multi-ion motion in the
bulk solutions is reduced to Coulomb interaction with ions arriving
at the channel mouth. The later process can be viewed as  a
stochastic Poisson process or as shot noise that modulates the
potential barrier for the conducting ion at the selectivity site.
The goal of the present paper is to estimate analytically the effect
of this potential modulation on the channel conductivity as will be
discussed in details in the next section.

To quantify the effect of multi-ion motion in the bulk on the
conducting ion at the selectivity site, we have simulated ion's
Brownian dynamics in the bulk. The resulting ion-ion distributions
in the bulk are shown in the Fig.~\ref{fig:distribution} (left). We
emphasize that these distributions are very close to those obtained
earlier in both BD simulations~\cite{Corry:01a} and
experiments~\cite{Guardia:91b}. The arrival time distributions for
Na$^+$ and Cl$^{-}$ ions at the channel mouth (defined as a
cylindrical section of radius $R$ and length $R$) obtained in our
simulations is shown in Fig.~\ref{fig:distribution} (right). Note
that these distributions  are exponential for both $Cl^{-}$ and
$Na^{+}$ with mean arrival times $\tau^{-}=471$ ps and
$\tau^{+}=365$ ps respectively. These estimates are in agreement
with the theoretical estimates in~\cite{Eisenberg:95}
\begin{equation}
\tau_{arrival} = \frac{1}{2\pi c D R}
\end{equation}
where  $c$ is the ion concentration and $D=\frac{k_BT}{m\gamma}$ is
the diffusion coefficient.

\begin{figure}[!h]
\begin{center}
\includegraphics[width=2.8in, height=2.5in]{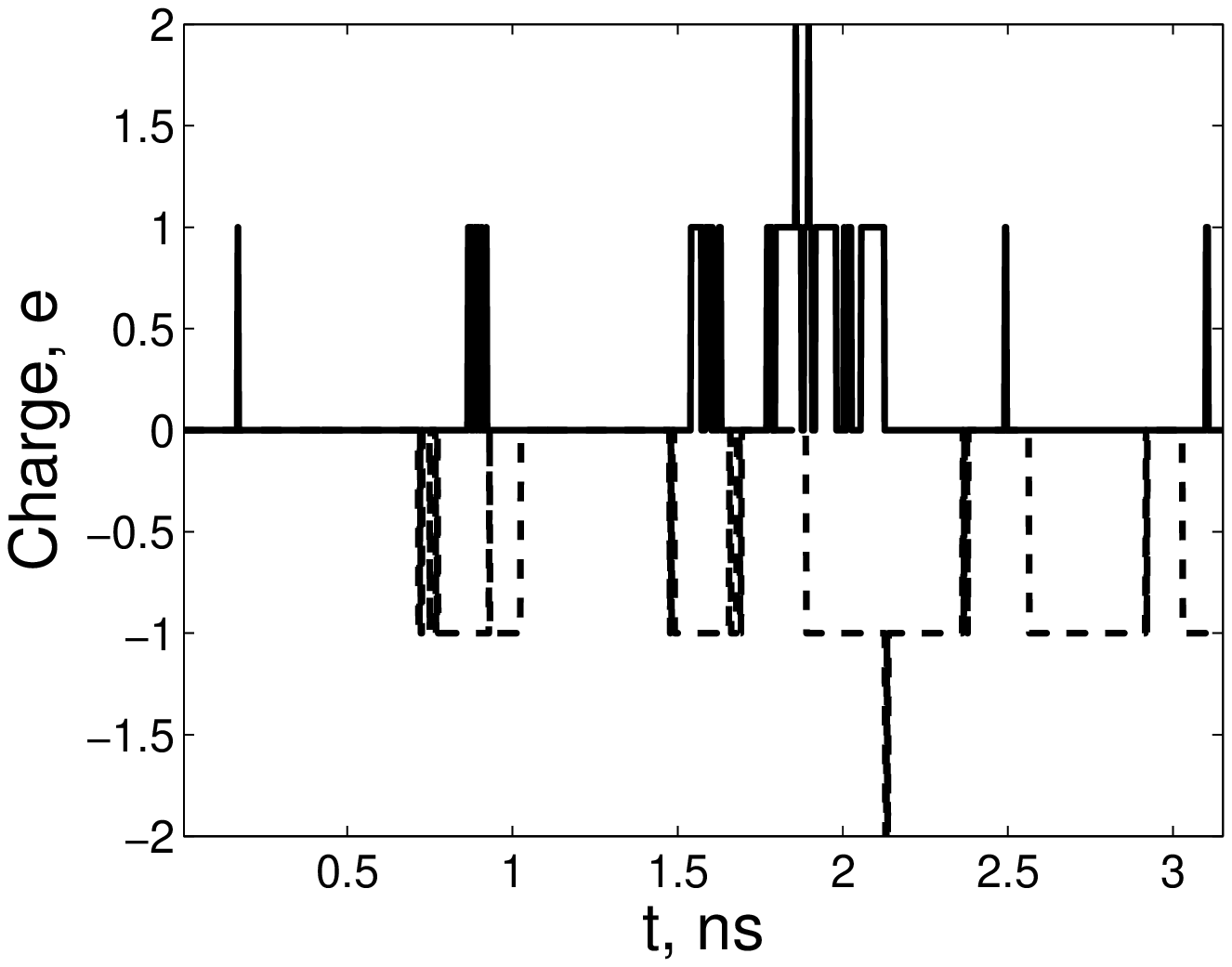}
\includegraphics[width=2.8in, height=2.5in]{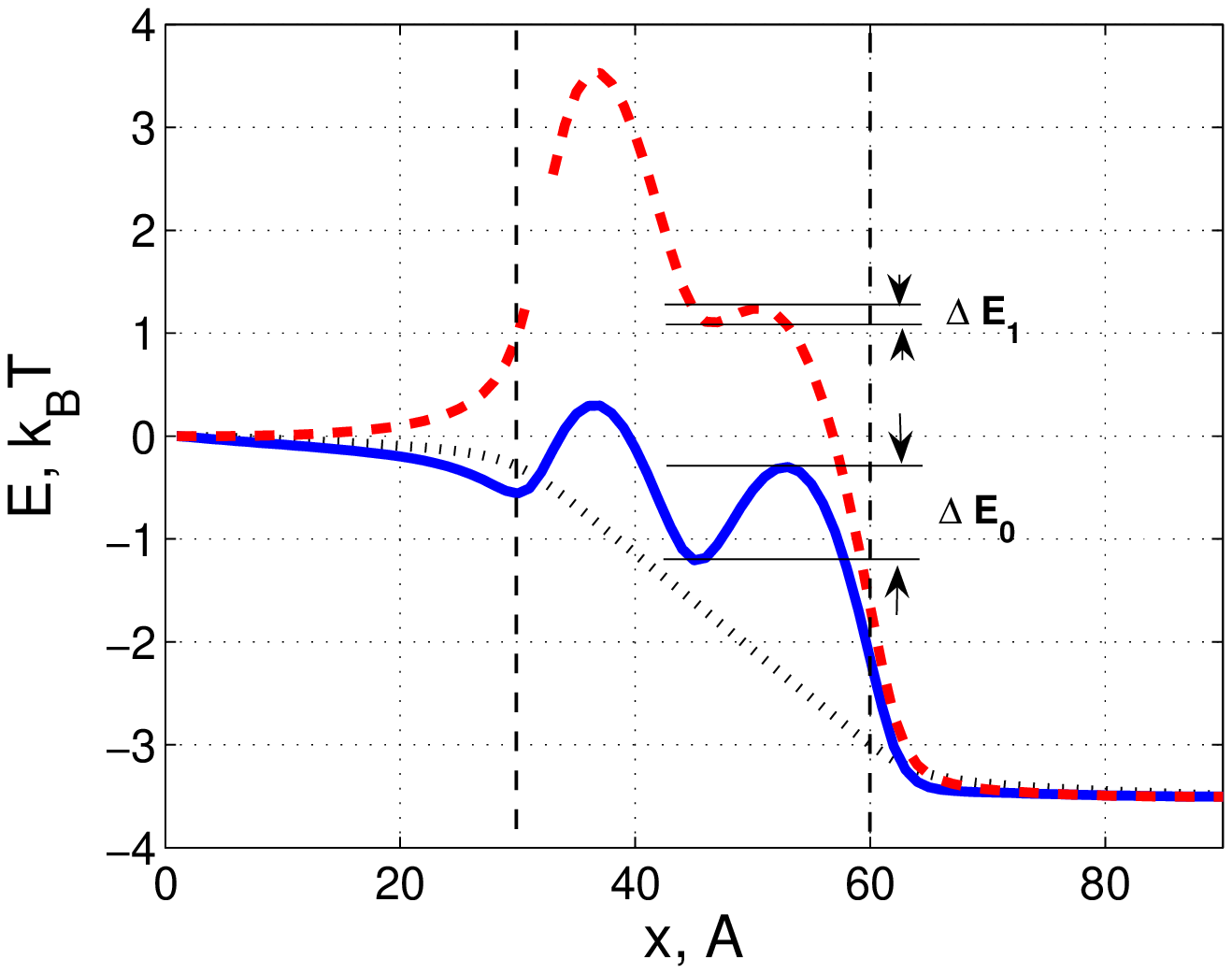}
\caption{\label{fig:energy} (left) Charge fluctuations at the
channel mouth. The positive charge is shown by the solid line. The
negative charge is shown by the dashed line. (right) The potential
energy profiles as a function of the position of the ion when: the
first ion is fixed at the channel mouth (dashed line) and the second
is moving along the channel axis. The solid line corresponds to the
potential energy on a single ion moving on the channel axis, and the
potential energy of the passive channel (dotted line). The vertical
dashed lines show the channel entrance and exit. The height of the
potential energy barrier seen by a single ion at the selectivity
site as it moves from left hand to right hand of the channel is
denoted $\Delta E_0$. In the presence of a second ion at the
channel's left mouth this barrier is reduced to $\Delta E_1$.}
\end{center}
\end{figure}

The time evolution of the charge in the channel mouth is shown in
Fig.~\ref{fig:energy} (left). It can be seen that the charge at the
channel mouth is a Poisson process with the three main states $+1e$,
$0$, and $-1e$. As a first approximation it is convenient to divide
the states of the channel potential affected by the charge
fluctuations into two effective states: (i) state of high
conductivity, corresponding to $+1e$, and (ii) state of low
conductivity, corresponding to $0$ or $-1e$ charge at the channel
mouth. In this approximation the effect of three states of the
potential is taken into account by asymmetry of the transition rates
between the two effective states. The corresponding transition rates
can be estimated as $\alpha^{\pm} = 1/\left<T_{\pm}\right>$, where
$\left<T_{\pm}\right>$ are mean residence time of two effective
states, giving $(\alpha^{+})^{-1} = 0.22$\,ns and $(\alpha^{-})^{-1}
= 0.38$\,ns. The occupation probabilities of these two states are
0.36 and 0.64 respectively. The effect of the wall oscillations on
the channel potential was estimated earlier~\cite{Luchinsky:07} and
for simplicity the wall will be assumed rigid in the rest of the
paper.

To estimate the effect of charge fluctuations on the value of the channel
potential we solve the Poisson equation for various positions of the conducting
ion along the channel axis in two cases: (i) when there are no other ions at
the channel entrances; (ii) when there is one positive ion at the left entrance
to the channel. The results of these calculations are shown in
Figure.~\ref{fig:energy}(right). The low-conductivity effective state of the
channel is shown by the blue solid line and corresponds to the potential
barrier $\approx 1k_BT$ at the selectivity site. The high conducting state is
shown by the red dashed line and corresponds to a potential barrier height
$\approx 0.2k_BT$, i.e.\ practically no barrier state. It can be seen that the
charge fluctuations are enhanced in channels of low dielectric constant,
resulting in strong modulation of the potential barrier at the selectivity
site~\cite{Luchinsky:07} (see also~\cite{Bastug:03b}).

\begin{figure}[!h]
\begin{center}
\includegraphics[width=3in, height=2.2in]{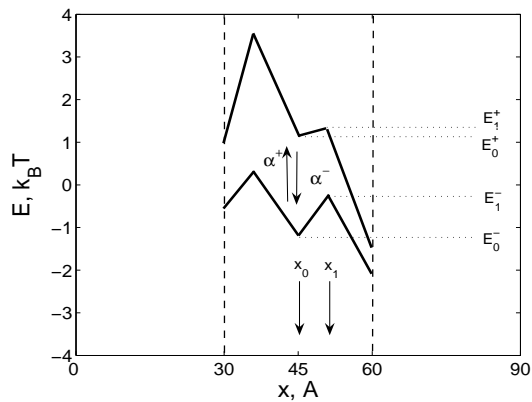}
\caption{\label{fig:fluctuating potential} Approximation of the
fluctuating potential.}
\end{center}
\end{figure}

It is therefore possible to build a simple model capable of coupling the motion
of ions in the channel to the bath solution. The channel potential becomes
\begin{eqnarray}
\label{eq:piecewise potential}
 &&V(x,t) = \frac{V_{+}+V_{-}}{2}+
\frac{V_{+}-V_{-}}{2}\chi(t), \quad \chi(t) = \pm1, \\
    &&V_{+}=\frac{\Delta E_{1}}{x_{m}}(x-x_{0})+E_{0}^{+}, \qquad
    V_{-}=\frac{\Delta E_{0}}{x_{m}}(x-x_{0})+E_{0}^{-}, \nonumber
\end{eqnarray}
where $\chi(t)$ is a Poisson random force with two transition rates
$\alpha_{\pm}$ between the states $+1$ and $-1$ . The charge fluctuations at
the channel mouth thus result in flipping of the potential. Here $\Delta
E_{0}=E_{1}^{-}-E_{0}^{-}$ and $\Delta E_{1}=E_{1}^{+}-E_{0}^{+}$, with $\Delta
E_{0}>\Delta E_{1}$ are respective barriers of the potential in two states, and
$x_{m}=x_{1}-x_{0}$.

A direct analogy can be made between the model described by
Eq.~(\ref{eq:overdamped ion}) and the model described by
Z\"{u}rcher~\cite{Doering:93a} whose barrier fluctuation is controlled by a
dichotomic noise of zero mean and exponential correlation. An approximation of
the fluctuating potential is sketched in Figure.~\ref{fig:fluctuating
potential}. The similarity of the two problems suggests that there is some
possibility of semi-analytical estimations of the effect of the charge
fluctuations.

\section{Estimation of the mean channel crossing time}
\label{s:crossing_time}

We are interested in the transition of a particle initially trapped at the
channel selectivity filter. This corresponds to the motion of the ion across
the interval $[x_{0},x_{1}]$. The approximate potential in this interval is
given by (\ref{eq:piecewise potential}). We are interested at the
unidirectional current, so there is no backward flow of ions. We assume that on
average, the channel is always occupied by a single Na$^{+}$ ion. This arises
from the fact that, when an ion is sitting in the middle of the channel, it is
almost impossible for a second ion to enter the channel due to the height of
the barrier at the left entrance of the channel, as can be seem from
Figure.~\ref{fig:fluctuating potential}.

As a first approximation, we assume that the mean first passage time (MFPT) is
only determined by escape. Therefore the MFPT for the channel is expected to be
a function of the two times $\tau_{+}$ and $\tau_{-}$ corresponding to the
escape times from the potential minimum in two effective states of the
potential. Our derivation follows very closely the earlier discussion by
Z\"{u}rcher~\cite{Doering:93a} with the difference that, here, we are
interested in the asymmetric case with two transitions rates.

Assuming no back flow, the quantities $\tau_{\pm}(x)$ are defined by
(see~\cite{Hanggi:85a}):
\begin{eqnarray}
\label{eq:equa1}
&&\frac{-1}{m\gamma}\frac{dV_{+}}{dx}\frac{d\tau_{+}}{dx}+\frac{k_{B}T}{m\gamma}\frac{d^{2}\tau_{+}}{dx^{2}}-\alpha^{+}
\tau_{+}+\alpha^{-}
\tau_{-}=-p_{+},\\
&&\frac{-1}{m\gamma}\frac{dV_{-}}{dx}\frac{d\tau_{-}}{dx}+\frac{k_{B}T}{m\gamma}\frac{d^{2}\tau_{-}}{dx^{2}}-\alpha^{-}
\tau_{-}+\alpha^{+} \tau_{+}=-p_{-},
\end{eqnarray}
The potential jumps between positive and negative values, with respective rates
$\alpha^{+}$ and $\alpha^{-}$. $p_{\pm}$ are the occupation probabilities of
these states. We choose a reflecting boundary condition (BC) at the bottom of
the barrier $x=x_{0}$ and absorbing BC at the top of the barrier $x=x_{1}$,
\begin{eqnarray}
\label{eq:equa3} \frac{d\tau_{\pm}(x=x_{0})}{dx}=0, \quad
\tau_{\pm}(x=x_{1})=0,
\end{eqnarray}
With $\tau_{+}$ and $\tau_{-}$ specified, the mean exit time of the
Brownian particle that is trapped at the selectivity filter
$x=x_{0}$ is given by:
\begin{eqnarray}
\label{eq:equa5} \tau = \tau_{+}(x_{0})+\tau_{-}(x_{0}),
\end{eqnarray}
Following Z\"{u}rcher~\cite{Doering:93a}, the calculations of
$\tau_{\pm}$ is straightforward. We introduce
\begin{eqnarray}
\label{eq:equa6} E = \frac{\Delta E_{1}+\Delta E_{0}}{2}, \quad
\Delta = \frac{\Delta E_{1}-\Delta E_{0}}{2},
\end{eqnarray}
and
\begin{eqnarray}
\label{eq:equa8} \sigma(x) =
\alpha^{+}\tau_{+}(x)+\alpha^{-}\tau_{-}(x), \quad \delta(x) =
\alpha^{+}\tau_{+}(x)-\alpha^{-}\tau_{-}(x).
\end{eqnarray}
We find the coupled differential equations
\begin{eqnarray}
\label{eq:equa10} \fl \frac{-DE}{x_{m}k_B
T}\frac{d\sigma}{dx}+D\frac{d^{2}\sigma}{dx^{2}}+(\alpha^{+}p_{+}+\alpha^{-}p_{-})=(\alpha^{+}p_{+}-\alpha^{-}p_{-})\delta
+ \frac{D\Delta}{x_{m}k_{B}T}\frac{d\delta}{dx},
\end{eqnarray}
\begin{eqnarray}
\label{eq:equa11} \fl \frac{-DE}{x_{m}k_B
T}\frac{d\delta}{dx}+D\frac{d^{2}\delta}{dx^{2}}-(\alpha^{+}p_{+}+\alpha^{-}p_{-})\delta
+(\alpha^{+}p_{+}-\alpha^{-}p_{-}) =
\frac{D\Delta}{x_{m}k_{B}T}\frac{d\sigma}{dx},
\end{eqnarray}
where $D=D_{Na}$ and the boundary conditions,
\begin{eqnarray}
\label{eq:equa12} \sigma(x=x_{1})=0,\;\;
\frac{d\sigma(x=x_{0})}{dx}=0, \quad \delta(x=x_{1})=0,\;\;
\frac{d\delta(x=x_{0})}{dx}=0.
\end{eqnarray}
We introduce the following scaled dimensionless coordinate,
\begin{eqnarray}
\label{eq:equa14} x = \frac{x_{m}k_{B}T}{E}y.
\end{eqnarray}
The model is therefore characterized by the following parameters,
\begin{eqnarray}
\label{eq:equa15}\fl \tau_{0} = 2\frac{(x_{m}k_{B}T)^{2}}{D E^{2}},
\quad \eta = \frac{\Delta}{E}, \quad \lambda =
\frac{\tau_{0}}{2}(\alpha^{+}p_{+}+\alpha^{-}p_{-}), \quad \beta =
\frac{\tau_{0}}{2}(\alpha^{+}p_{+}-\alpha^{-}p_{-}).
\end{eqnarray}
The coupled differential equations then read,
\begin{eqnarray}
\label{eq:equa19}
-\frac{d\sigma}{dy}+\frac{d^{2}\sigma}{dy^{2}}+\lambda=\beta\delta+\eta\frac{d\delta}{dy},
\end{eqnarray}
\begin{eqnarray}
\label{eq:equa20}
-\frac{d\delta}{dy}+\frac{d^{2}\delta}{dy^{2}}-\lambda\delta + \beta
= \eta\frac{d\sigma}{dy},
\end{eqnarray}

%---------------------
\begin{figure}[!t]
\begin{center}
\includegraphics[width=2.8in, height=2.2in]{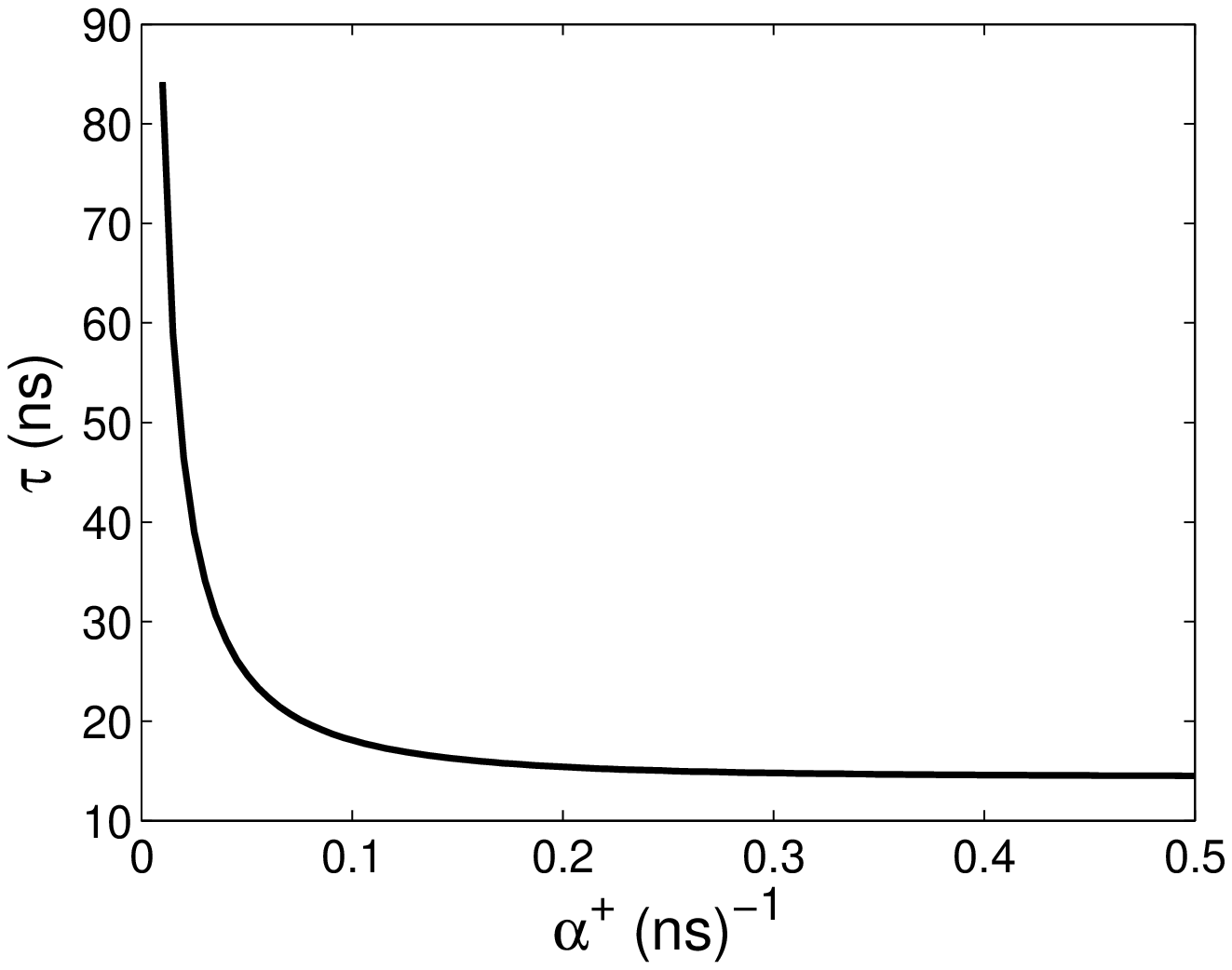}
\includegraphics[width=2.8in, height=2.2in]{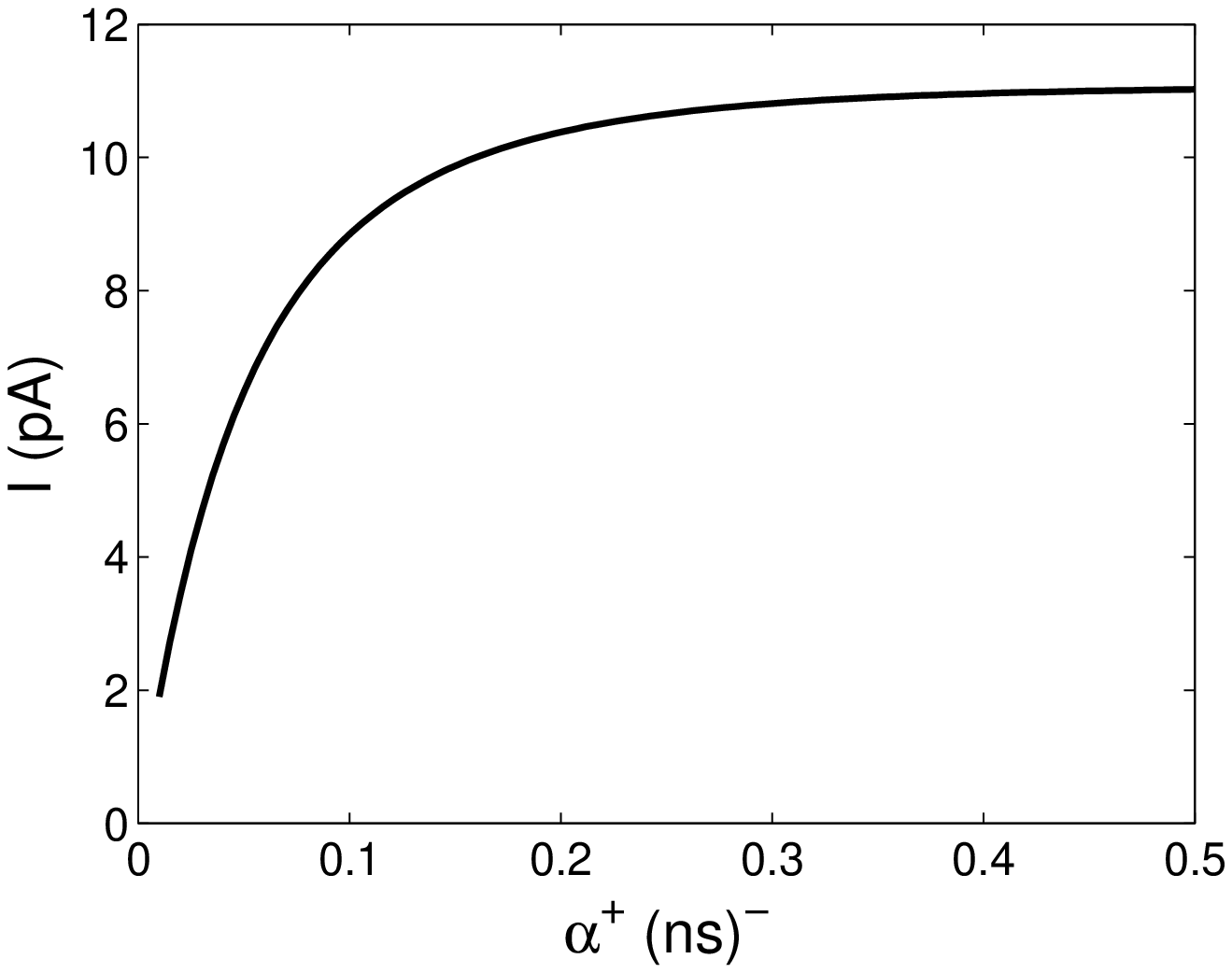}
\caption{\label{fig:tau_alpha1} {\bf Left:} MFPT as a function of
$\alpha^{+}$. {\bf Right:} Current as a function of $\alpha^{+}$.}
\end{center}
\end{figure}
%--------------------

\noindent and $\sigma(y)$ and $\delta(y)$ are subject to the BC,
\begin{eqnarray}
\label{eq:equa21} \sigma(y=y_{1})=0,\;\;
\frac{d\sigma(y=y_{0})}{dy}=0, \quad \delta(y=y_{1})=0,\;\;
\frac{d\delta(y=y_{0})}{dy}=0.
\end{eqnarray}
The solution of this system gives,
\begin{eqnarray}
\label{eq:equa23} \delta(y) =
\sum_{i=1}^{3}a_{i}\exp(q_{i}y)+\frac{(\beta-\lambda\eta)}{(\lambda-\beta\eta)}
\end{eqnarray}
The eigenvalues $q_{i}$, follow from
\begin{eqnarray}
\label{eq:equa24} \left\{
  \begin{array}{ll}
    &q_{1}+q_{2}+q_{3}=2, \\
    &q_{1}q_{2}+q_{2}q_{3}+q_{3}q_{1}=1-\lambda-\eta^{2},\\
    &q_{1}q_{2}q_{3}=-(\lambda-\beta\eta).
  \end{array}
\right.
\end{eqnarray}
The $a_{i}$ are given as follows,
\begin{eqnarray}
\label{eq:equa25}
a_{1} = -\frac{\beta-\lambda\eta}{D_{s}(\lambda-\beta\eta)}\bigg(q_{3}(q_{2}^{2}-\lambda)-q_{2}(q_{3}^{2}-\lambda) \bigg)\exp((q_{3}+q_{2})y_{0})\nonumber \\
-\frac{1}{D_{s}}\bigg(\frac{\lambda(\beta-\lambda\eta)}{\lambda-\beta\eta}-\beta
\bigg)\bigg(q_{3}\exp(q_{3}y_{0}+q_{2}y_{1})-q_{2}\exp(q_{3}y_{1}+q_{2}y_{0})
\bigg).
\end{eqnarray}
Here $a_{2}$ and $a_{3}$ are determined by cyclic permutation of the indices of
the $q_{i}$ from $a_{1}$, and $D_{s}$ is given by
\begin{eqnarray}
\label{eq:equa26} D_{s} =
q_{1}q_{2}(q_{2}-q_{1})(q_{3}-1)\exp((q_{1}+q_{2})y_{0}+q_{3}y_{1})
+ cycl. perm.
\end{eqnarray}
At $y_{0}$, we have:
\begin{eqnarray}
\label{eq:equa27}
\sigma(y_{0}) = \frac{1}{\eta}\sum_{i=1}^{3}a_{i}\bigg[\bigg(-1+q_{i}-\frac{\lambda}{q_{i}}\bigg)\exp(q_{i}y_{0})-\bigg(q_{i}-\frac{\lambda}{q_{i}}\bigg)\exp(q_{i}y_{1}) \bigg]\nonumber \\
-\frac{(\beta-\lambda\eta)}{\eta(\lambda-\beta\eta)}+\bigg[\frac{\lambda}{\eta}\frac{(\beta-\lambda\eta)}{(\lambda-\beta\eta)}
-\frac{\beta}{\eta}\bigg](y_{1}-y_{0})
\end{eqnarray}
Combining Eq.~(\ref{eq:equa5},\ref{eq:equa8}), the mean exit time
for the Brownian particle follows:
\begin{eqnarray}
\label{eq:equa28} \tau =
\frac{\sigma(y_{0})+\delta(y_{0})}{2\alpha^{+}}+\frac{\sigma(y_{0})-\delta(y_{0})}{2\alpha^{-}}.
\end{eqnarray}

%---------------------
\begin{figure}[!h]
\begin{center}
\includegraphics[width=2.8in, height=2.5in]{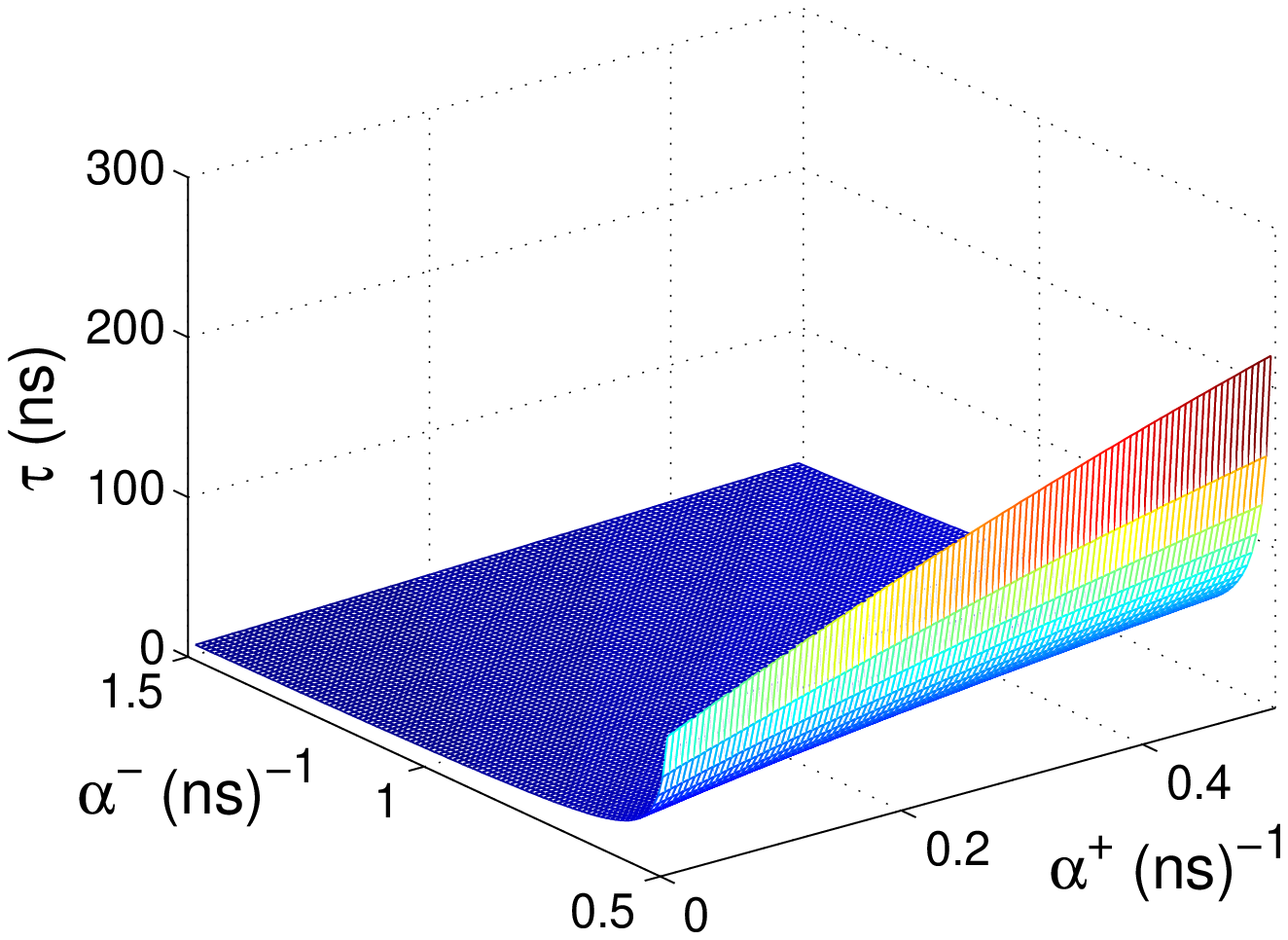}
\includegraphics[width=2.8in, height=2.5in]{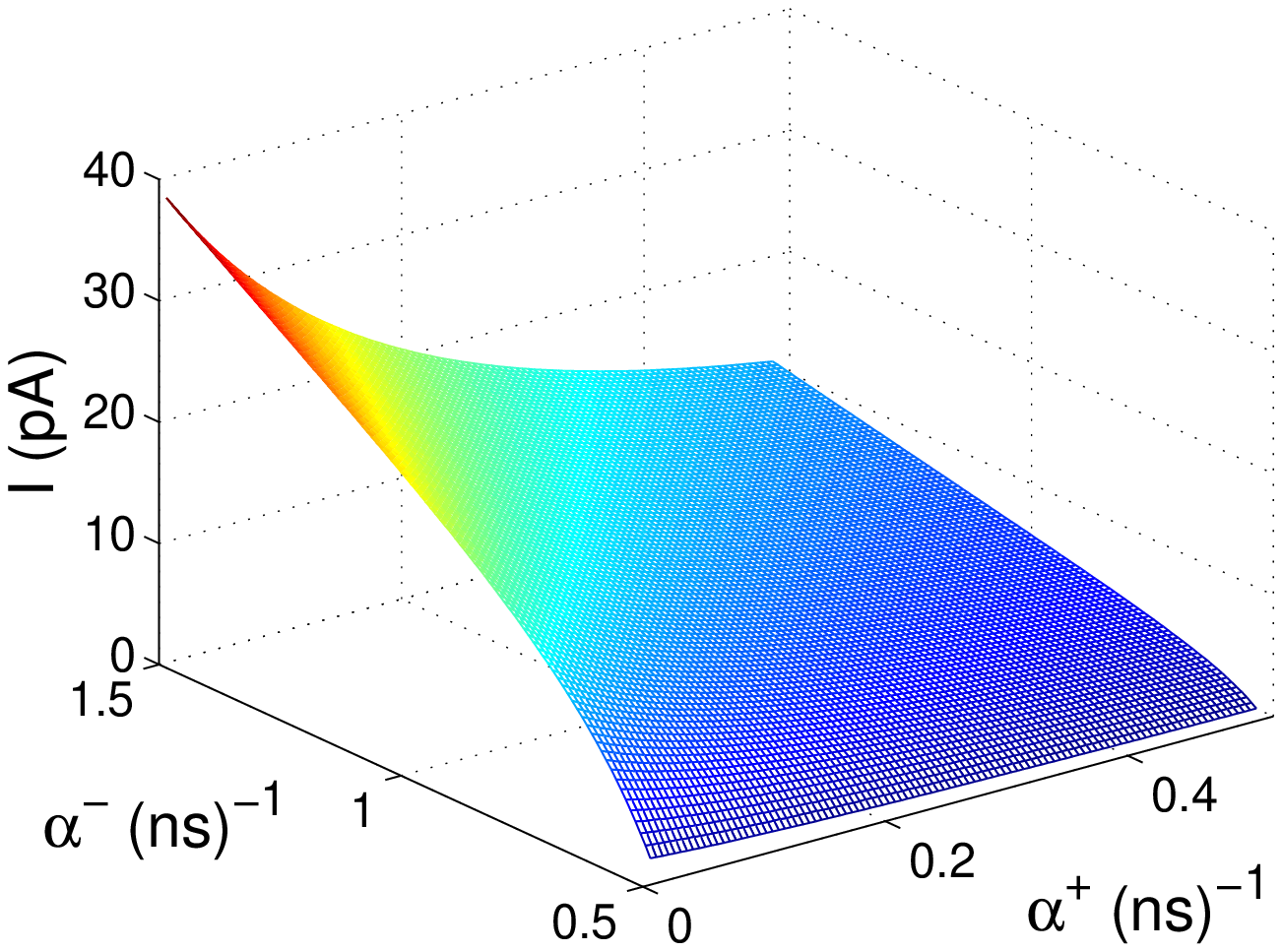}
\caption{\label{fig:tau_alpha2} (left) MFPT as function of
$\alpha^{+}$ and $\alpha^{-}$. (right) current as function of $\alpha^{+}$ and $\alpha^{-}$.}
\end{center}
\end{figure}
%--------------------

The effect of shot noise on ion channel conduction is related to the
arrival of ions at the channel's mouth. We calculated the channel
MFPT as a function of the two flipping rates. Since the channel has
a high affinity for the ions and therefore slow them down, we have
used a smaller ion diffusion coefficient inside the channel equal to
$1.33\times10^{-10}$ m$^{2}$s$^{-1}$. The results of the
calculations are shown in the Figs.~\ref{fig:tau_alpha1}
and~\ref{fig:tau_alpha2}. It is clear from Fig.~\ref{fig:tau_alpha1}
that the particle takes longer to cross $\Delta E_{0}$ as compared
to $\Delta E_{1}$. There is a fast drop of the MFPT as the flipping
rate $\alpha^{+}$ increase. The current $I=e/\tau$ is also presented
as function of $\alpha^{+}$.  A more general view of the dependance
of the MFPT and the current on the two rates is shown in
Fig.~\ref{fig:tau_alpha2}. We emphasize that the obtained transition
rates are essentially non-equilibrium. In particular, current
saturation effect can be observed as $\alpha^{+}$ increases in a
wide range of parameters. This is in accordance with experimental
observations~\cite{Andersen:05} of the current saturation at high
concentrations.

\section{Conclusion}
\label{s:conclusions}

In summary, we have introduced a Brownian dynamical model of ionic
transitions through a channel, taking into account charge
fluctuations at the channel mouth and the fluctuations of the
channel walls. The statistical properties of the charge fluctuations
are reconstructed from 3D Brownian dynamics simulation of multi-ion
motion in the bulk solution. It is shown that distributions of ion
arrival times at the channel mouth are exponential. It is further
shown that these charge fluctuations strongly modulate the potential
barrier for the conducting ion at the selectivity site due to
amplification of electrostatic interactions in long narrow channels
of low dielectric constant. These findings have allowed us to model
the mean ion transition time through the channel as an ionic escape
from the potential wall at the selectivity site induced by thermal
fluctuations and by modulation of the height of the barrier by
stochastic Poisson processes. The derived model is a Brownian
dynamical model of the ``knock-on'' mechanism of the
type~\cite{Hodgkin:55a,Berneche:03a}. Our model allows for analytic
estimation of transition probabilities in the presence of charge
fluctuations, i.e.\ it allows for analytic estimation of
correlations between bulk concentrations and ion currents in charged
narrow channels. In particular, it demonstrates the effect of
current saturation due to ion concentrations in the bath. The model
is essentially of a non-equilibrium nature. This last point is worth
emphasizing because traditional approaches assume equilibrium rates
of hopping between the sites.

We note that the model takes into account the wall fluctuations. The
latter feature is very important e.g.\ for an analysis of the
tightly correlated motion at selectivity site of the type discovered
in KcsA~\cite{Kutluay:05a}. Our model allows for analytic estimation
of transition rates in the presence of oscillations of channel
walls, using our earlier results~\cite{Luchinsky:00o} on escape from
periodically driven potentials using the method of logarithmic
susceptibility~\cite{Dykman:97b} as will be discussed in details
elsewhere. Such an escape process, assisted by the periodic
modulation of the potential barrier by the wall oscillations, can
result on its own in selectivity between alike ions due to the
difference in their diffusion
coefficients~\cite{Marchesoni:98a,Sintes:02a}. However, ultimately
the selectivity of the channel has to be incorporated into the model
by taking into account the effects of hydration~\cite{Choe:98a} (see
also~\cite{Luchinsky:07} for a discussion of how hydration effect
can be incorporated in our model). It is also worth mentioning that
the model that takes into account fluctuations of the wall may
account for the dissipation of energy in the channel, for
self-induced acceleration of the transition rate of the ion through
the channel, and for coupling of the ion motion to the channel
gating mechanics. Indeed, in this model, part of the energy induced
by a very strong Coulomb interaction between the charged site at the
channel wall and the moving ion is stored as the energy of
vibrational modes. The later energy is only partially dissipated by
the protein phonon modes. The remaining energy can now be used to
modulate the potential barrier for the moving ion in a
self-consistent manner to accelerate its transition through the
channel. It can also be used to assist the conformational changes
leading to the channel gating.

The work in progress contains a plethora of unsolved problems. The
immediate extension of the proposed model will be to include more
then two levels for the potential at the selectivity site, taking
into account positive, neutral, and negative charge at the channel
mouths at both ends of the channel. The model can be further refined
by including the estimates of the return times corresponding to a
return of the ion in the channel to the initial bulk solution.

Following the discussion above we can also formulate the following general
unsolved problems in ion channels. We believe that they can be tackled by
extending the model introduced in this paper:
\begin{itemize}
\item[(i)] The role of the membrane fluctuations.
\item[(ii)] The role of the hydration potential.
\item[(iii)] The role of additional binding sites outside the selectivity
    filter.
\item[(iv)] The energetics of the ion transition including energy
    relaxation due to the coupling to the protein phonon modes (wall
    oscillations).
\item[(v)] The coupling of the ion-wall interaction to the gating
    mechanism.
\end{itemize}

\noindent In all of these, noise and dynamical effects seem to play a crucial
role that is only starting to be elucidated. Our preliminary research shows
that the model can be extended to take all of these effects into account.

\section*{References}

%\bibliographystyle{iopart-num}
%\bibliography{my_thesis,channels,channels2,joint}

\providecommand{\newblock}{}

\end{document}